\documentstyle[preprint,aps]{revtex}
\begin{document}

\draft

\title{Influence of gauge fluctuations on fermion pairing order parameter}

\author{Yu-Liang Liu}
\address{Center for Advanced Study, Tsinghua University, Beijing 100084, 
People's Republic of China}

\maketitle

\begin{abstract}

Using a prototype model, we study the influence of gauge fluctuations on
fermion pairing order parameter which has the gauge symmetry, and demonstrate
that the gauge fluctuations can destroy the long range order of the fermion
pairing order parameter, and make it only have short range correlation. If 
this parameter is a superconducting order parameter, we show that the Meissner
effect of the system keeps intact, and the system is in the superconducting 
state even though the long range order of the superconducting order parameter
is destroied by the gauge fluctuations. Our calculations support that
the pseudo-gap region of the high Tc cuprate superconductivity is a spin
pseudo-gap region rather than an electron pre-paired region.

\end{abstract}
\vspace{1cm}


\newpage

After the discovery of the high Tc cuprate superconductor\cite{1}, it has been
soon reached a common consensus that the high Tc cuprate superconductors are a 
two-dimensional strongly correlated system, and low energy behavior is 
determined by their copper-oxide plane(s)\cite{2,2',3,4}. 
However, so far a complete theory representing this system does not appear,
and there are still a lot of controversies and unsolved questions in this
field. This strongly correlated electron system has novel physical properties
of normal and superconducting states. In normal state, the
system shows non-Fermi liquid behavior\cite{2',5,6,7}, 
and has strong antiferromagnetic fluctuations in low doping and optimal
doping regions. Moreover, in the low doping region
there is a pseudo-gap region, and stripe phase fluctuations may also appear.
These phenomena cannot be clearly explained by an unified theory. 
In superconducting state, the system has a d-wave symmetry superconducting
order parameter, but it is not clear which mechanism can produce this order
parameter. There is another puzzle question that single electron excitation
spectrum keeps intact as the system going to the superconducting state
from the pseudo-gap region\cite{8,8',9}. This phenomenon seems to mean that 
the pseudo-gap region is an electron pre-paired region\cite{10}.

Usual perturbation methods are hard to treat strongly correlated systems,
because in which there is not a controllable small parameter, and 
electrons have strong correlation. For enough strong electron correlation,
there are not well-defined quasi-particles near the Fermi surface, and the 
systems show non-Fermi liquid behavior. On the other hand,
these perturbation methods 
cannot directly describe the electron correlation. However, the electron
correlation is a key parameter hidden in the strongly correlated systems.
The eigen-functional bosonization method\cite{11} is a good candidate for 
treating the strongly correlated systems, because in this method there
naturally appears a phase field, where 
its imaginary part represents the electron correlation.

In this paper, we introduce
a prototype model, which represents a two-dimensional spin-1/2 fermion system
with transverse gauge fields, to study the influence of the gauge fluctuations
on the fermion pairing order (spin pseudo-gap) parameter, here we assume
that the fermion pairing order parameter does not violate the gauge symmetry, 
and these transverse gauge fields originate from some other strong fermion
interactions, such as the Hubbard model with large on-site electron
Coulomb interaction, or the electron single occupied constraint of the 
t-J model. In the absence of the gauge fields, if the fermion pairing
parameter is a constant (long range order), we demonstrate that after turning
on the gauge fields, the correlation function and the average value of the
fermion pairing order parameter go to zero with anomalous exponential 
asymptotic behavior. The transverse gauge fields provide a strong phase
fluctuation factor to the fermion pairing parameter by phase fields. If these
fermions are electrons, we demonstrate that the Meissner effect of the system
keeps intact, and the system is still in the superconducting state, even
though the long range order of the superconducting order parameter is destroied
by the transverse gauge fields. This result is similar to that of a 
two-dimensional superconductor with long range Coulomb interaction, where
the long range order of the superconducting parameter is destroied by the
long range Coulomb interaction, but the Meissner effect of the system leaves
intact\cite{12}. However, there is a little difference between these two cases,
for the long range Coulomb interaction, the correlation function of the
electron pairing order parameter shows power-law asymptotic behavior.

We consider a prototype model represented by
the Hamiltonian with transverse gauge fields ${\bf A}(x)$
(${\bf \nabla}\cdot{\bf A}(x)=0$),
\begin{eqnarray}
H &=& \displaystyle{ \int d^{2}x \left\{ \psi^{\dagger}_{1}(x)[\frac{1}{2m}
(-i{\bf \nabla}-g{\bf A})^{2}-\mu]\psi_{1}(x)\right.} \nonumber \\
&+& \displaystyle{ \psi^{\dagger}_{2}(x)[\frac{1}{2m}(-i{\bf \nabla}+
g{\bf A})^{2}-\mu]\psi_{2}(x)} \label{1} \\
&+& \displaystyle{ \left.\Delta(x)\psi_{1}(x)\psi_{2}(x)+\Delta^{*}(x)
\psi^{\dagger}_{2}(x)\psi^{\dagger}_{1}(x)\right\}}
\nonumber\end{eqnarray}
where the fermion field $\psi_{1}(x)$ represents the fermions with spin-up,
the fermion field $\psi_{2}(x)$ represents the fermions with spin-down, and 
$\Delta(x)=-u<\psi^{\dagger}_{2}(x)\psi^{\dagger}_{1}(x)>$ is fermion
pairing order parameter (spin pseudo-gap)
induced by other interactions, where $u$ is the interaction strength. 
It is noted that the fermion pairing terms
do not violate the gauge symmetry of the transverse gauge fields. 
At ${\bf A}(x)=0$, for simplifying
our following calculations, we simply take this parameter as a real
constant, $\Delta(x)|_{{\bf A}=0}=\Delta_{0}$. The system described by the
Hamiltonian (\ref{1}) is a strongly correlated system, as $\Delta(x)=0$ it
shows the non-Fermi liquid behavior\cite{13,14}.

With the Hamiltonian (\ref{1}), 
the action of the system can be written as a simple form,
\begin{equation}
S=\int dt d^{2}x \Psi^{\dagger}(x,t)\hat{M}(x,t)\Psi(x,t)
\label{2}\end{equation}
where $\Psi^{\dagger}(x,t)=(\psi^{\dagger}_{1}(x,t),\psi_{2}(x,t))$, $\hat{M}
=\hat{M}_{0}+\hat{\phi}(x,t)$ and 
\begin{eqnarray}
\hat{M}_{0} &=& \displaystyle{\left( \begin{array}{c} \displaystyle{
i\partial_{t}+\mu+\frac{\nabla^{2}}{2m}, \;\;\;
\Delta_{0}}\\  \Delta_{0}, \;\;\; \displaystyle{
i\partial_{t}-\mu-\frac{\nabla^{2}}{2m}}
\end{array}\right)} \nonumber \\
\hat{\phi}(x,t) &=& \displaystyle{\left( \begin{array}{c} \displaystyle{
-\frac{ig}{m}{\bf A}(x,t)\cdot{\bf \nabla}-\frac{g^{2}}{2m}{\bf A}^{2}(x,t)},
\;\;\; \Delta^{*}(x,t)-\Delta_{0}\\  \Delta(x,t)-\Delta_{0}, \;\;\;
\displaystyle{ \frac{ig}{m}{\bf A}(x,t)\cdot{\bf \nabla}+\frac{g^{2}}{2m}
{\bf A}^{2}(x,t)}\end{array}\right)}
\nonumber\end{eqnarray}
We have separated the propagator operator $\hat{M}(x,t)$ into two parts, one
is a free term, and another one depends on the transverse gauge fields 
${\bf A}(x,t)$. In this form, we first exactly solve the eigen-equation of 
the operator $\hat{M}_{0}$, then we determine the differential equations of
phase fields induced by the transverse gauge fields.

Integrating out the fermion fields $\psi_{1}(x,t)$ and $\psi_{2}(x,t)$, we
obtain the effective action (omitting constant term),
\begin{equation}
S[A]=-i\int^{1}_{0}d\xi\int dt d^{2}x tr\left(\hat{\phi}(x,t)G(x,t;x',t',[\xi
\phi])\right)|_{\stackrel{t'\rightarrow t}{x'\rightarrow x}}
\label{3}\end{equation}
where the Green's functional $G(x,t;x',t',[\phi])$ can be represented by 
eigen-functionals of the propagator operator $\hat{M}(x,t)$,
\begin{equation}
G(x,t;x',t',[\phi])=\sum_{i}\sum_{k,\omega}\frac{1}{E^{(i)}_{k\omega}[\phi]}
\Psi^{(i)}_{k\omega}(x,t,[\phi])\Psi^{(i)*}_{k\omega}(x',t',[\phi])
\label{4}\end{equation}
The eigen-functional equation of the operator $\hat{M}(x,t)$ reads,
\begin{equation}
\hat{M}(x,t)\Psi^{(i)}_{k\omega}(x,t,[\phi])=E^{(i)}_{k\omega}[\phi]
\Psi^{(i)}_{k\omega}(x,t,[\phi])
\label{5}\end{equation}
where $i=1,2$ and 
the eigenvalues are $E^{(1,2)}_{k\omega}[\phi]=\omega\pm E_{k}+
\Sigma^{(1,2)}_{k}[\phi]$, where $E_{k}=\sqrt{\epsilon^{2}(k)+
\Delta^{2}_{0}}$, $\epsilon(k)=k^{2}/(2m)-\mu$, and $\Sigma^{(i)}_{k}[\phi]=
\int^{1}_{0}d\xi\int dt d^{2}x\Psi^{(i)*}_{k\omega}(x,t,[\phi])\hat{\phi}(x,t)
\Psi^{(i)}_{k\omega}(x,t,[\phi])$ is a smooth function, and independent of
$\omega$. Due to the 
non-zero fermion pairing parameter $\Delta_{0}$, the energy
spectrum of the fermions splits into two branches, one represents unoccupied
states (conducting band), and another one represents occupied states (valence
band). The fermion pairs are in the top of the valence band. The 
eigen-functionals $\Psi^{(1.2)}_{k\omega}(x,t,[\phi])$ represent these 
occupied and unoccupied states, respectively, and they can be written as,
\begin{eqnarray}
\Psi^{(1)}_{k\omega}(x,t,[\phi]) &=& \displaystyle{ A_{k}\left(\frac{1}{
TL^{2}}\right)^{1/2}\left(\begin{array}{c}
v_{k}e^{Q_{k}(x,t)}\\ u_{k}e^{\tilde{Q}_{k}(x,t)}\end{array}\right)
e^{i{\bf k}\cdot{\bf x}-i(\omega+\Sigma^{(1)}_{k}[\phi])t}} \nonumber \\
\Psi^{(2)}_{k\omega}(x,t,[\phi]) &=& \displaystyle{ A_{k}\left(\frac{1}{
TL^{2}}\right)^{1/2}\left(\begin{array}{c}
u_{k}e^{F_{k}(x,t)}\\ -v_{k}e^{\tilde{F}_{k}(x,t)}\end{array}\right)
e^{i{\bf k}\cdot{\bf x}-i(\omega+\Sigma^{(2)}_{k}[\phi])t}}
\label{6}\end{eqnarray}
where $u_{k}=\sqrt{(1+
\epsilon(k)/E_{k})/2}$ and $v_{k}=\sqrt{(1-\epsilon(k)/E_{k})/2}$ are
coherence factors,
$A_{k}$ is a normalization constant, and $T$ and $L$ are the time and space
length scales of the system, respectively. At ${\bf A}(x,t)=0$, the 
eigen-functionals in (\ref{6}) are the exact solutions of the operator
$\hat{M}_{0}$, where $A_{k}=1$.
The fermion pairing parameter now
can be approximately written as near the Fermi surface ($k\sim k_{F}$),
\begin{equation}
\Delta(x,t)=\Delta_{0}e^{\tilde{Q}_{k}(x,t)+Q^{*}_{k}(x,t)}
\label{7}\end{equation}
where $\Delta_{0}$ is determined by the self-consistent equation
$\Delta_{0}=\frac{u}{2L^{2}}\sum^{'}_{k}|A_{k}|^{2}\Delta_{0}/E_{k}$. The phase
fields $Q_{k}(x,t)$ and $\tilde{Q}_{k}(x,t)$ satisfy Eikonal-type 
equations\cite{11}. If we only keep linear terms, we can obtain that,
\begin{equation}
\left\{ \begin{array}{ll}
\displaystyle{ [i\partial_{t}+i\frac{{\bf k}\cdot{\bf \nabla}}{m}+
\frac{\nabla^{2}}{2m}]Q_{k}(x,t)+\frac{g}{m}{\bf k}\cdot{\bf A}(x,t)}= &
-\Delta_{0}[\tilde{Q}_{k}(x,t)+\tilde{Q}^{*}_{k}(x,t)]\\
\displaystyle{ [i\partial_{t}-i\frac{{\bf k}\cdot{\bf \nabla}}{m}-
\frac{\nabla^{2}}{2m}]\tilde{Q}_{k}(x,t)-\frac{g}{m}{\bf k}\cdot{\bf A}(x,t)}
= & -\Delta_{0}[Q_{k}(x,t)+Q^{*}_{k}(x,t)] \end{array}\right.
\label{8}\end{equation}
Under this approximation, we have the same effective action as that 
obtained by usual
random-phase approximation (RPA) (see below). 
By solving these equations, we have the relations between the phase fields and 
the transverse gauge fields, 
\begin{equation}
\left\{ \begin{array}{c} \displaystyle{ Q^{R}_{k}(q,\Omega)=
-\frac{gq^{2}}{m^{2}D_{-}}{\bf k}\cdot{\bf A}(q,\Omega)}\\  \displaystyle{
\tilde{Q}^{R}_{k}(q,\Omega)=-\frac{gq^{2}}{m^{2}D_{+}}{\bf k}\cdot{\bf A}
(q,\Omega)} \end{array}\right., \;\;\;\;
\left\{ \begin{array}{c} \displaystyle{ Q^{I}_{k}(q,\Omega)=
-\frac{2g(\Omega-\frac{{\bf k}\cdot{\bf q}}{m})}{mD_{-}}{\bf k}\cdot{\bf A}
(q,\Omega)}\\  \displaystyle{ \tilde{Q}^{I}_{k}(q,\Omega)=
\frac{2g(\Omega+\frac{{\bf k}\cdot{\bf q}}{m})}{mD_{+}}{\bf k}\cdot{\bf A}
(q,\Omega)}\end{array}\right.
\label{9}\end{equation}
where $D_{\pm}=(\Omega\pm{\bf k}\cdot{\bf q}/m)^{2}-q^{4}/(2m)^{2}\mp
\Delta_{0}q^{2}/m$, $Q_{k}(q,\Omega)=Q^{R}_{k}(q,\Omega)+Q^{I}_{k}(q,\Omega)$,
and $\tilde{Q}_{k}(q,\Omega)=\tilde{Q}^{R}_{k}(q,\Omega)+\tilde{Q}^{I}_{k}
(q,\Omega)$. The imaginary parts of the phase fields do not contribute to 
the effective action, but they represent the fermion correlation induced by the
transverse gauge fields. Only the real parts of the phase fields contribute
to the effective action of the system.

With equations (\ref{4}) and (\ref{9}), we have the
effective action,
\begin{equation}
S[A]=\frac{1}{TL^{2}}\sum_{q,\Omega}\Pi_{ij}(q,\Omega)A_{i}(-q,-\Omega)
A_{j}(q,\Omega)
\label{10}\end{equation}
where $\Pi_{ij}(q,\Omega)=(i\Gamma(\Delta_{0})\Omega/q+\chi(\Delta_{0})
q^{2})(\delta_{ij}-q_{i}q_{j}/q^{2})$, and $\Gamma(\Delta_{0})=
\gamma(\Delta_{0})+2m\Delta_{0}(\gamma+\gamma(\Delta_{0}))/q^{2}$. The 
parameters $\gamma(\Delta_{0})$ and $\chi(\Delta_{0})$ are the smooth functions
of the gap $\Delta_{0}$. At $\Delta_{0}=0$, we have $\gamma(0)=\gamma$ and
$\chi(0)=\chi$, and have the same effective action of the transverse gauge
fields ${\bf A}(x,t)$ as that of Refs.\cite{5,6}, 
where it was obtained by usual
RPA method. We would like to point out that in the eigen-functional
bosonization method it is naturally to introduce the phase field, which is
a key parameter hidden in strongly correlated systems, because its imaginary
part represents the fermion correlation. However, it is absent in previous
perturbation methods. With the phase field, we can easily study the influence
of fermion interactions on the low energy behavior of the system, and 
calculate a variety of correlation functions.

At $\Delta_{0}=0$, the system shows non-Fermi liquid behavior produced by 
the strong fluctuations of the transverse gauge fields\cite{13,14}. It can
be easily shown by present representation. With equation (\ref{4}) and taking
functional average of the transverse gauge fields ${\bf A}(x,t)$, we can
obtain single fermion Green's function,
\begin{eqnarray}
G_{1}(x-x', t-t') &=& i<G_{11}(x,t;x',t',[\phi])>_{{\bf A}} \nonumber \\
&=& \displaystyle{ \frac{1}{L^{2}}\sum_{k}\theta(-\epsilon(k))
e^{i{\bf k}\cdot({\bf x}-{\bf x}^{'})-i\epsilon(k)(t-t')}
e^{P^{I}_{k}(x-x',t-t')}}
\label{11}\end{eqnarray}
where $P^{I}_{k}(x,t)$ comes from the contribution of the imaginary part of
the phase field $Q_{k}(x,t)$, and the contribution from the real part of the 
phase field $Q_{k}(x,t)$ can be neglected. Near the Fermi surface $k\sim
k_{F}$, we have the relation,
\begin{equation}
P^{I}_{k_{F}}(0,t)\simeq -i^{1/3}ag^{2}t^{1/3}
\label{12}\end{equation}
where $a=\frac{k_{F}}{4\pi^{2}m}(\chi/\gamma)^{1/3}\int dx(1-\cos(x))x^{-4/3}$.
This result is basical same as that of Ref.\cite{13,14}. 
Due to the time dependence 
of the phase factor $P^{I}_{k_{F}}(0,t)$, the single fermion Green's function
shows a singular low energy dependence which violates the quasi-particle
excitation assumption of the Landau Fermi liquid. Therefore, the system shows
non-Fermi liquid behavior.

We now study the influence of gauge fluctuations on the fermion pairing order
parameter. From
equation (\ref{7}), we see that the gauge fields ${\bf A}(x,t)$ provide a
phase fluctuation factor
to the fermion pairing parameter $\Delta(x,t)$ by the phase 
fields $Q_{k}(x,t)$ and $\tilde{Q}_{k}(x,t)$. Therefore, the gauge fields
strongly affect the fermion pairing parameter. 
By taking functional average of the
transverse gauge fields, we have the following relations,
\begin{eqnarray}
<\Delta(x,t)\Delta^{*}(x,t)>_{{\bf A}} &\simeq & \Delta^{2}_{0}, \;\;\;\;
<\Delta(x,t)\Delta^{*}(x',t)>_{{\bf A}}\simeq \displaystyle{
\Delta^{2}_{0}e^{-b|{\bf x}-{\bf x'}|}}, \nonumber \\
<\Delta(x,t)\Delta^{*}(x,t')>_{{\bf A}} &\sim & \displaystyle{
e^{-c|t-t'|^{1/5}}}, \;\;\;\;
<\Delta(x,t)>_{{\bf A}}\sim \displaystyle{ e^{-d/\omega^{1/5}_{0}}}, 
\label{13}\end{eqnarray}
where $b\sim \Delta^{-1/5}_{0}$,
$c\sim i^{1/5}\frac{k_{F}}{4\pi^{2}m}[\chi(\Delta_{0})/(2m\Delta_{0}
(\gamma+\gamma(\Delta_{0})))]^{1/5}$, $2d\sim c$ is a constant, 
and $\omega_{0}$ is 
a characteristic energy scale of the transverse gauge fields, in general it is
very small ($\omega_{0}\sim 1/L$). These relations are our central results, we 
now explain them. The first expression means that the real parts of the
phase fields $Q_{k}(x,t)$ and $\tilde{Q}_{k}(x,t)$ have little influence on
the spin pseudo-gap, and the last three relations mean that the imaginary
parts of the phase fields strongly suppress the correlation of the fermion
pairing parameter $\Delta(x,t)$, and make the long range order of the fermion
pairing parameter be a short range order. Therefore, the fluctuations of
the transverse gauge fields can destroy the long range order of the fermion
pairing parameter, and make it only have short range correlation.

If these fermions are electrons, the pairing parameter $\Delta(x,t)$ is a
superconducting order parameter. At ${\bf A}(x,t)=0$, the Hamiltonian (\ref{1})
represents a two-dimensional superconductor with superconducting gap
$\Delta_{0}$. The transverse gauge fluctuations make the system have a short
range pairing parameter $\Delta(x,t)$, but they do not destroy the 
superconducting state of the system. This can be justified by calculating the
Meissner effect of the system after turning on an external magnetic field
$B(x,t)={\bf \nabla}\times{\bf A}_{e}(x,t)$. The current functional reads,
\begin{eqnarray}
{\bf J}(x,t,[A]) &=& \displaystyle{ -\psi^{\dagger}_{1}(x,t)\frac{i{\bf 
\nabla}}{m}\psi_{1}(x,t)-\psi^{\dagger}_{2}(x,t)\frac{i{\bf \nabla}}{m}
\psi_{2}(x,t)} \label{14} \\
&-& \displaystyle{ \frac{1}{m}[({\bf A}_{e}(x,t)+g{\bf A}(x,t))
\psi^{\dagger}_{1}(x,t)\psi_{1}(x,t)+({\bf A}_{e}(x,t)-g{\bf A}(x,t))
\psi^{\dagger}_{2}(x,t)\psi_{2}(x,t)]}
\nonumber\end{eqnarray}
Taking functional average of the transverse gauge fields, we have the relation,
\begin{eqnarray}
{\bf J}(x,t) &=& <{\bf J}(x,t,[A])>_{{\bf A}} \nonumber \\
&=& \displaystyle{ \left(\frac{a(\Delta_{0})}{m^{2}}-\frac{\rho(\Delta_{0})}
{m}\right){\bf A}_{e}(x,t)}
\label{15}\end{eqnarray}
where $\rho(\Delta_{0})=-(1/L^{2})\sum_{k}\theta(-\epsilon(k))\epsilon(k)
/E_{k}$, and $a(\Delta_{0})=(1/L^{2})\sum_{k}\frac{k^{2}}{2E^{2}_{k}}
\frac{E_{k}E_{k'}-\epsilon(k)\epsilon(k')}{E_{k}+E_{k'}}|_{{\bf k}^{'}
\rightarrow{\bf k}}$. At $\Delta_{0}=0$, we have ${\bf J}(x,t)=0$. This 
equation shows that even though the transverse gauge fields destroy the
long range superconducting order, the Meissner effect of the system keeps
intact, i.e., the system still is in the superconducting state.
This result is consistent with that of equation (\ref{13}). The Meissner effect
is mainly determined by the coherence factors $u_{k}$ and $v_{k}$ of the
superconducting state, and is insensitive to the phase factor of the electron
pairing order parameter. However, the fluctuations of the transverse gauge 
fields can destroy the long range order of the electron pairing order 
parameter. One must remember that these gauge fields are not electromagnetic
gauge fields, they originate from some strong electron interactions, and the
superconducting state of the system does not violate this gauge symmetry.

Even though all our calculations are based on the s-wave symmetry of the
pairing parameter, they are also qualitatively valid for the d-wave symmetry
of the pairing parameter.
The above results may provide some useful informations for studying the 
high Tc cuprate superconductors, where their low energy physical properties
are determined by the two-dimensional strongly correlated electron system in 
the copper-oxide plane(s). It is well-known that this strongly correlated 
system can be well described by the t-J model\cite{3} or the Hubbard model
with large on-site Coulomb interaction\cite{2}. Electron single occupation
constraint of the t-J model in fact represents the strong electron correlation,
which can induce the gauge interaction among the fermions and holons in the
slave boson/fermion representation of the electron operator\cite{15,16,16'}, 
or the gauge interaction of the electrons due to the Hilbert space of the 
system is strongly suppressed by this constraint. We now only focus on the
pseudo-gap region in the low doping region. If this region is an electron
pre-paired region, and the change of electron pairing order 
parameter from long range order to short range order
is induced by the gauge fluctuations, our calculations show that
under an external magnetic field, the Meissner effect of the system will
be observed, and the system is in the superconducting state,
which contradicts with present experimental data. In fact, the system is in
a normal state, and there is not the Meissner effect in this region. However,
our calculations support that the pseudo-gap region is a spin pseudo-gap 
region, and the fermions do not have electric charges. The transport property
of the normal state is determined by the holons (carrying electric charges).
Due to strong gauge fluctuations, the system still shows non-Fermi liquid
behavior even though there is the spin pseudo-gap in spin excitation
spectrum. This is qualitatively consistent with present experimental 
observations. How does the system go into the superconducting state?
There may have two ways, one is usual Bose-Einstein condensation of the 
holons\cite{6,15}, however, in this case there is too high onset temperature
of superfluidity. Another one is that in the low temperature limit the gauge
fluctuations become enough strong to re-combine a fermion and a holon into 
an electron, and the fermion pairing order parameter becomes the
superconducting order parameter. In this case, the superconducting 
order parameter has the same symmetry (d-wave symmetry) and the same
modulus with the fermion pairing order parameter,
which is qualitatively consistent with the angle-resolved 
photoemission experiments\cite{8,8',9}, 
where single electron excitation spectrum
nearly has no change as the system going to the superconducting state
from the spin pseudo-gap region.

In summary, with the prototype model, we have studied the influence of the
transverse gauge fields on the fermion pairing order parameter which keeps
the gauge symmetry, and demonstrated that the gauge fluctuations destroy the
long range order of the fermion pairing order parameter, and make it only 
have the short range correlation. If this order parameter is the 
superconducting order parameter, we demonstrated that the Meissner effect of 
the system keeps intact, and the system is in the superconducting state even
though the long range order of the superconducting order parameter is
destroied by the gauge fluctuations. Therefore, our calculations support that
the pseudo-gap region in the low doping region of the high Tc cuprate
superconductivity is the spin pseudo-gap region rather than the electron
pre-paired region, because the high Tc cuprate superconductivity is a strongly
correlated electron system, where the Hilbert space of the electron states is
strongly suppressed, the gauge fluctuations are existing.

\end{document}